\pdfoutput=1
\documentclass[crop]{CUPLStd}
\jname{International Journal of Astrobiology}%
\makeatletter
\providecommand{\@LN}[2]{}
\makeatother

\usepackage[authoryear]{natbib}

\usepackage{amsmath} 
\usepackage{graphics} 
\usepackage{subfig}

\begin{document}

\supertitle{Research Paper}

\title[]{On possible   life-dispersal patterns  beyond the Earth}

\author[]{Andjelka Kova{\v c}evi{\'c}$^{1,2}$}

\address{\add{1}{Department of astronomy, Faculty of Mathematics, University of Belgrade, Studentski trg 16, 11000 Belgrade, Serbia} \\
\add{2}{Fellow of Chinese Academy of Sciences President's International Fellowship Initiative (PIFI) for visiting scientist}}

\corres{\name{A. Kova{\v c}evi{\'c}} \email{andjelka@matf.bg.ac.rs}}

\begin{abstract}

The assumption that exoplanets are {' in equilibrium'} with their surroundings has not given way to life's transmissivity on large spatial scales.
The spread of human diseases and the  life recovery {rate} after mass extinctions on our planet, {on the other hand}, may exhibit spatial and temporal scaling as well as distribution correlations that influence the mappable range of their characteristics.
We model hypothetical bio-dispersal within {a} single Galactic region using the stochastic infection dynamics process, which is inspired by these local properties of life dispersal on Earth.
We split the population of stellar systems into different categories regarding habitability and evolved them through time using probabilistic cellular automata rules analogous to the model.
As {a} dynamic effect, we include  {the} existence of natural dispersal vectors (e.g., dust, asteroids)  in a way that avoids assumptions about their agency (i.e., questions of existence). {Moreover}, by assuming that dispersal vectors have a finite velocity and range, the model includes {the} parameter of 'optical depth of life spreading'.
The effect of the oscillatory infection rate ($b(t,d)$) on the long-term behavior of the dispersal flux, which adds a diffusive component to its progression, is also taken into account. The life recovery rate ($g(t,d)$) was only included in the model as a link to macrofaunal diversity data, which shows that all mass extinctions have a 10-Myr ‘speed rate' in diversity recovery. {This parameter accounts} for the repopulation of empty viable niches as well as the formation of new {ones}, {without ruling out} the possibility of {genuine} life reemergence on other habitable worlds in the Galaxy that {colossal} extinctions {have} sterilized.
All life-transmission events within the Galactic patch have thus been mapped into phase space characterized by parameters $b$ and $g$.
We found that phase space is separated into subregions of long-lasting transmission, rapidly terminated transmission, and a transition region between the two.
We observed that depending on the amplitude of the oscillatory life spreading rate, life-transmission in the Galactic patch might take on different geometrical shapes (i.e., 'waves'). 
Even {if} some host systems are uninhabited,  life transmission has a certain threshold, allowing a patch to be saturated with viable material over a long period.
Although stochastic fluctuations in {the} local density of habitable systems allow for clusters that can continuously infect one another, the spatial pattern disappears when life transmission is below the observed threshold, {so that} transmission process is not permanent in time.
Both findings {suggest} that a habitable planet in a densely populated region {may} remain uninfected.

\end{abstract}


\selfcitation{}

\received{xx xxxx xxxx}

\revised{xx xxxx xxxx}

\accepted{xx xxxx xxxx}

\maketitle
\Fpagebreak

\section{Introduction}

{
{Astrobiology has several dissonances}, but the most well-known is between the probability and evidence of extraterrestrial life.}
{Detectability} of extraterrestrial life is {linked to} the likelihood of its (un)correlated origin.
{ Due to a} synergy of physical limitations of life forms, the velocity of natural vehicles (solar and interstellar meteorites, comets), and physical conditions in interstellar space, the probability {of observing} correlated origin (i.e., dispersal origin) of extraterrestrial life decreases, {even though} the chance of finding {it} increases with the distance from the Earth.
{Abiogenesis,} the spontaneous process {through which} non-viable matter becomes live organisms, {appears to be the most plausible} life origin mechanism on planets based on these {factors} \citep{10.1007/s11214-010-9671-x}.

{
Some conceptual studies ponder if standard, limiting definitions of life will prevent future astrobiological missions from identifying extraterrestrial life, and propose alternate definitions \citep{10.1089/ast.2010.0524,10.3390/life10040042}.
In this light, we will summarize the basic intuitions \citep{10.3390/life10040042} employed in our research rather than attempting to establish even a rudimentary definition of life: 
 (1) Terrestrial life as we know it may be uncommon throughout the universe, but a universal class of entities with life-like attributes may be far more prevalent (2) There may be entities that match the living characteristics better than even earthly life does, but they have yet to be discovered or even considered; (3) We can unravel knowledge and investigate the complete parameter space of physical and chemical interactions that may produce and propagate life by loosening limits on the notion of life.}

\indent 
{Scientific} evidence of life and its (un)correlated genesis, {on the other hand}, is expected to have mappable distributions {throughout} a wide range of spatial dimensions, from microscopic ($<$mm) to macroscopic ($>$AU) levels. 
{Only} probabilistic models, statistical representations that produce dependencies that allow inference and prediction between scales, can be used to examine this range of astrobiological scales \citep[see][]{10.1017/S1473550410000042, 10.1089/ast.2017.1782, 10.1089/ast.2018.1903}.

{ Although biology shares some length-scale characteristics with chemistry and physics, it is distinct in that it is a dynamic science, as organisms and their lineages change dramatically over time. DNA transcription/replication can take anywhere from minutes to hours ($10^{-6}- 10^{-4}$yr), whereas evolutionary change can take hundreds to thousands ($10^{3} - 10^{5}$) of years, affecting organisms lineage.}

{
In the context of life spreading on Galactic scales, cellular automata (CA) simulations are particularly noteworthy. In astrobiology, the use of probabilistic cellular automata (PCA) models was pioneered by
 \citet{10.1007/s11084-012-9293-2}, and followed by \citet{10.1017/S1473550412000420}. \citet{10.1017/S1473550418000101}, recently,  built  a model based  on an earlier study by \citet{cond-mat/0112137}.  \citet{10.1051/0004-6361/201834588} have also explored the relationship between spreading life and catastrophic processes in a similar discrete model.}

\indent 
{The Galactic habitable zone (GHZ) and the circumstellar habitable zone (CHZ), {postulated by} \citet{10.1006/icar.2001.6617} and \citet{10.1126/science.1092322} {and formalized} by \citet{10.1006/icar.1993.1010} distinguish regions in our Galaxy {primarily} on their levels of habitability.}
{These ground-breaking studies paved the way for important discussions of correlations involved across the scales.} {Within these habitable zones,} the spatial distribution of stars, exoplanets, and exomoons differentiates these worlds according to {the extent} of habitability. {When} \cite{10.1089/ast.2013.1088} {considered} which worlds might be more habitable than Earth,  an interesting point {went ahead}. The authors propose that, just as the Solar System {proved} to be an outlier among planetary systems, Earth may prove to be an outlier among habitable or, eventually, inhabited worlds.  {As a result}, {it is} possible that Earth is a planet with sub-optimal habitability {compared} to other superhabitable worlds. At {the} very least, compared to the {equivalent} regions of other stars, {the solar} CHZ may be a suboptimal location in our Galaxy. 
We {still} have {a lot} to learn about habitability, but similar to biogeography, we might{ be able to find} clusters of habitable zones in some parts of our Galaxy where life {can thrive}.
Any network of habitability zones that is eventually mapped {may show} dispersal routes or linkages between types of life that appear at nodes in the network. Such network analysis, for example, has been successfully utilized to investigate hydrothermal vent biogeography, as described by  \cite{10.1093/sysbio/syr088}.
Biogeography {uses} geographical and dispersion routes to explain species distribution. Gradual range dispersal, {for example}, is the continued transmission of populations by a series of conventional (short-distance) dispersal events, typically over {extended} time scales incorporating evolutionary processes. { Jumping dispersal occurs} when organisms overcome a barrier in a single, direct action rather than bouncing across 'islands.'.
Sweepstakes routes {are formed} when organisms overcome a barrier {unusually and dangerously}, with just a few individuals surviving the journey \citep{10.1016/B0-12-226865-2/00073-0}.

\indent {Astrobiology} may now examine some of the {possible} dispersal vectors and routes { within} the solar system {thanks to advancements} in space technology. \citep[see][and references therein]{10.1007/978-981-13-3639-3_27}. 
{Life} spreading between planets in a single planetary system, such as the Solar System, would be considered as the short-range routes. Microscopic life could {migrate} between Earth and Mars, for example, via meteorites or comets, although this has {yet to be} verified.
\cite{10.1089/153110703321632525} {determined} that any rock {ejected} from a terrestrial planet in our Solar System will never collide with a similar terrestrial world in an exoplanetary system; in this view, short-range correlation effects are limited to the planetary system {in question} \citep{10.1073/pnas.1703517114}.  \citet{10.1111/j.1365-2966.2004.07287.x}, {on the other hand, hypothesized} that short-range correlation might have evolved into long-range correlation. It {happens} when meter-scale boulders {blasted} from Earth collide with other objects and dust particles in space, resulting in {remains} as small as a micron. These {remnant particles} can {then} be ejected from the Solar System as a result of radiation pressure.

\indent {The number of} observations of interstellar objects {traveling} through our system has recently increased  
\citep[e.g., Oumuamua asteroid and 2I$/$Borisov comet][]
{10.1038/s41550-017-0361-4, 10.3847/2041-8213/ab530b}.
{Interstellar objects like these} could {be used to} transport organic material between planetary systems. 
{However, no} microorganisms or fossils have yet been discovered in martian meteorites that have collided with Earth \citep{10.1126/science.278.5344.1706}.
{Nonetheless, some} research imply that life transmission is far more common in crowded environments. \citep[see][]{Bel12}.

{Future} and current space missions will {aid in} the investigation of these life-dispersal {paths} in our and exoplanetary systems, {as well} as in the regions {of} our Galaxy. If future investigations {discover} dispersal {channels} for life, it will have {a significant impact} on {how} future space life search initiatives {are interpreted}.
\indent \citet{10.1088/2041-8205/810/1/l3} {presented a novel} observational approach for {resolving this problem.}  
{ The authors  created a generic statistical strategy based on pandemic models on Earth  propose that discovering biologically active planets within particular galactic areas separated by regions where planetary life is rare or absent can be practically incontrovertible proof for life transmission.}

\indent \citet{10.1089/ast.2005.5.497}, {for example, examined } the probability of life exchange in star-forming clusters, where exoplanetary systems are closer{ together} and their relative speeds are lower.  \citet{10.1017/S1473550419000144} {has demonstrated} that interactions between clouds of small bodies {from approaching stellar} systems could play a role in interstellar life exchange. {Comets can be exchanged} between stars due to gravitational {disruption} of these clouds. {Furthermore,} in young star clusters, an exchange of planets between systems can occur during their close encounters, which could be another  {pathway} for interstellar life exchange \citep[see][]{10.1111/j.1365-2966.2010.17730.x}.

\cite{10.3847/2041-8213/aaef2d}  {suggested} that viruses,{in addition to} bacteria, may spread across the Milky Way.
{No biochemical derivatives such as methane or $\mathrm{N}_{2}\mathrm{O}$ should be detected remotely because viruses do not metabolize directly. However, understanding these alterations to Earth's biogeochemical cycles, the consequences of virus infection at an extraterrestrial ecosystem-level may be detectable. \citep[e.g.,][]{10.1089/ast.2017.1649}}.

\indent What we {do} know is that the high-profile infections diseases (such SARS, H5N1, HIV, Ebola, and COVID-19) {show} large-scale regional and temporal patterns, i.e. spatio-temporal clustering \citep{CO,10.1016/j.annepidem.2016.12.001, 10.1126/science.abb3221}.
{At least three} characteristics of interplanetary organic material transport and epidemic data {are challenges} to using traditional statistical methodologies. The data are rarely the result of planned experiments {in both domains}, the occurrences (observations) should not be independent, and the process is often only partially observable \citep{10.18637/jss.v077.i11}. Surveillance technology in the field of astrobiology, {however}, has not yet reached its full potential.

\indent 
Rather than {considering} unrelated events (where abiogenesis occurs with no relation to neighboring regions), we investigate what to expect from a correlated life origin due to a "contagion" hypothesis.
We report on numerical tests carried out in the framework of a 2-dimensional Cellular Automaton (CA) model for {the} local {Galactic} region.

{The} search for biosignatures with the European Extremely Large Telescope and space observatories (such as the James Webb Space Telescope \citep[JWST,][]{10.1007/s11214-006-8315-7} or Atmospheric Remote-sensing Exoplanet Large-survey \citet[ARIEL,][]{10.1007/s10686-018-9598-x}) will {initially} be limited to the local region of our star system (i.e., within a few tens of light-years) {due to} technical constraints. {Therefore}, the cellular automata model we present here concentrates on a straightforward life diffusion scenario within a small region of our Galaxy.

{We} aim to get a better understanding of life-diffusion within a galactic patch by taking into consideration the spatial density and viable material mobility process. {Furthermore}, we want to establish how to decode phase space thresholds {for extinction and persistence of life diffusion}.

\indent The following sections compose the content of this article. We discuss the statistical toy model for life spreading in Section Methods, which takes into account individual interactions {among hosts in the galaxy patch by capturing viable material}. {Then}, the section Results and Discussion analyzes the numerical experiments and displays the phase space of life spreading as well as its properties. The key conclusions are summarized in the final section.



\section{Methods}

Instead of CA models {considering} the entire Galaxy, the CA we present here primarily considers a simple life diffusion scenario {inside} a small patch because the { quest} for biosignatures will initially only { span} a few tens of light-years beyond our system.

{Here, we demonstrate how a SIR model  \citep[firstly introduced by ][]{10.1098/rspa.1927.0118}  implemented in a CA context can be used to simulate the dynamics of life diffusion.
Kermack and McKendrick's work has been cited numerous times and has become a classic in infectious disease epidemiology \citep{10.1080/17513758.2012.716454}. {The papers of} Kermack and McKendrick influenced {the creation of} mathematical models for disease spread in {the twentieth and twenty-first centuries}, and they are still applicable in many epidemic circumstances \citep{10.1016/j.physa.2020.125659}.}

CA  parameters have been {adjusted to account for Galactic patch characteristics while} remaining compatible with SIR pandemic models.  We {separated the population of host systems} into various habitability categories {and maintained} them through time using CA rules that {are equivalent} to non-linear ordinary differential equations (ODE):

\begin{align}
\dot{S}&=-b SI+\tau_{R}R;\\
\dot{I}&=b SI+\tau_{lf}I;\\
\dot{R}&=\tau_{lf} SI-\tau_{R}R.
\end{align}
\noindent {{The} terms $S$ and $I$, respectively, denote the number of habitable and inhabited hosts. $R$ is the total number of sterilized hosts. According to the first equation, an infected host ($I$) can infect habitable  ($S$) neighbors at a rate of $b$. The second equation shows that hosts live for a length of time$\tau_{lf}$ before becoming sterilized ($R$). At rate $\tau_R$, a sterilized host becomes habitable again, according to the third equation.  
The above equations are defined at the time $t$ and the spatial position $x$, such that $(t,x)\in [0,T]\times \Omega$, $\Omega$ is a fixed and bounded domain in $\mathbb{R}\times \mathbb{R}$ with smooth boundary $\partial \Omega$, and homogeneous Neumann boundary conditions $\frac{\partial S}{\partial \eta}=\frac{\partial I}{\partial \eta}=\frac{\partial R }{\partial \eta}=0$, where $\eta$ is the outward unit normal vector on the boundary. The homogeneous Neumann boundary condition implies that the system described above is self-contained, and no emigration occurs beyond the boundary. }

Our CA model for the Galactic patch includes {a} cluster of matter and voids, assuming Newtonian gravitational law. In a square lattice, we suppose that matter and emptiness are of equal size.
{CA cells} can represent individual hosts as well as planetary systems or different stars due to the universal nature of transition rules. {For the sake of simplicity}, we will refer to them as hosts.
As a result, the cell might represent a planetary system, a star, or             a void.
The terms planetary system and star do not always refer to the singular objects that these terms {typically} signify.

 CA is a 2D square lattice {with} $N\times N$ lattice cells and $\mathcal{T}$ total time steps,
  with non-periodic boundary conditions (i.e., it has the topology of 2D squared field).
{It} has matrix representation $CA^{l}\in R^{m\times m}$ at each time step, {with} elements  $ca^{l}_{ij}, \, l \in{1,\cdots, \mathcal{T}}, i,j \in{1,\cdots,m}$. 
Our cellular automata is a four-state model, { which means} that each cell {could only be} in one of the four possible states.
{A lattice cell state is
$ca^{k}_{ij} =0$ if there is a habitable host (i.e. planetary system) at the ith row and the 
jth column, or it is  $ca^{k}_{ij} =1$ if contains life, or it is  $ca^{k}_{ij} =2$  if host is  sterilized and finally it may be empty cell (void) $ca^{k}_{ij} =3$. 
{In a biological sense}, the first and last states are passive, the second is active, {whereas} the third is just physically active.} 
The number of neighbors, $n$, around a cell is calculated using the following equation:
$n=(2*r+1)^2$
where $r$ is the range.  {We use a} Moore Neighborhood {with a range of} $r=1$, which gives each cell eight possible neighbors.
{Further, we randomly} populate the cells with hosts in the Galactic patch's initial state using the axisymmetric ansatz for stellar number density $n$ \citep{10.3847/1538-3881/aa8ef1}:

\begin{equation}
\begin{aligned}\label{eq:quadratic}
n= c \cdot e^{-0.5\sqrt{X^{2}+Y^{2}+Z^2}},
\end{aligned}
\end{equation}
\noindent where $c\sim 1\, \mathrm{pc^{-3}}$,  $X,Y, Z$ are galactocentric coordinates but  for simplicity sake the third coordinate is taken as  $Z=0$.

{The state transition} for the cell at location $(i,j)$ at time step $k$ is as follows:
the habitable cell can be infected with life by an
infected neighbor with {an} infection rate $b(t,d)$, where $t$ and $d$ {represent} time and distance, respectively. This rule is a state transition $0\rightarrow 1$.
{The host will bear life {at a} random constant rate $\tau_{lf}\in Rnd\{0,0.05\}$,where $Rnd$ stands for randomly chosen value. A  (pseudo) random numbers are generated using the Fast Mersenne Twister SFMT19937 generator  \citep{10.1007/978-3-642-04107-5_38}, a high-quality  generator with an exceptionally large period  $2^{19937} - 1$ or about $10^{6001}$ of the produced sequence.}
A  life -bearing cell  can be sterilized  {by applying} the (immunity) rate  $g(t,d)$, which corresponds to a state transition $ 1\rightarrow2$.
We {also account for} the possibility of life {recovering} in the sterilized cell at varying rates $\tau_{lr}\in\{0.25,0.5\}$.
{A} state transition $2\rightarrow1$ corresponds to {this scenario}.
{
The life recovery rate was included in the model only as a comparison to findings from macrofaunal diversity data, which show that all mass extinctions have a 10-Myr 'speed rate' in diversity recovery \citep{10.1073/pnas.0802597105,10.1038/s41559-019-0835-0}.
  However, unlike diversity, morphological complexity can be rebounded more quickly, reaching a plateau within $\sim5$\,Myr following the Cretaceous mass extinction, as shown for planktic foraminifera \citep{10.1038/s41559-019-0835-0}.}
 { Furthermore,} we let a life-terminated host become habitable again at a rate of $\tau_{R }=0.2$, i.e., state transition $2\rightarrow0$.

{We regard the dispersion of matter} content (e.g., dust, asteroids) or the presence of additional dispersal vectors as dynamic effects.
{As a result,} we assumed that the infectious connections between CA cells {are determined by} a capture kernel: the rate {at} which viable material {emanating} from the host at location $l 1$ {is} captured at position $l 2$. The kernel can only be defined up to a proportionality constant, and we {opt} to  {have it} in Gaussian form:

\begin{equation}
\begin{aligned}\label{kappa}
\kappa(d)=p_{0}exp^{-d^2/2\sigma^2},
\end{aligned}
\end{equation}

\noindent where $d$ is the Euclidean distance between the cells.
{We choose} $p_{0}=1$ for the parameter $ p_0$, which is infection probability at zero distance (i.e., when an infected cell is coincident with a habitable).
'Optical depth of life spreading' is controlled by the scale parameter $\sigma$, which {governs the pace} at which probability declines with distance. 
{We let the} parameter  $\sigma$ to be 50.
{Also, we assume} that the temporal {component} of $b(t,d)$ is periodic throughout time in our model:
\begin{equation}
\begin{aligned}\label{beta}
b(t,d)=\beta \cdot (1+\sin\frac{2\pi t}{	P}*\kappa(d)),
\end{aligned}
\end{equation}

\noindent where $\beta$ is a variable parameter, $t$ is a time, and $P$ is {the} infection period. 
{
We arbitrarily chose  $P\sim 10\%\mathcal{T}$ as a compromise between our ignorance and the {possibility} of substantial variability in such processes.}
For the immunity rate, we applied the following formula:

\begin{equation}
\begin{aligned}\label{gamma}
g(d)=g_{0}\kappa(d),
\end{aligned}
\end{equation}

\noindent where  $g_0$ {represents} a variable sterilizing rate.
{We set} transmission of life for all events to be proportional to a parameter $b(t,d)$ defining infection rate because both rate kernels $b(t,d)$ and $g(d)$ are proportional to $kappa(d)$. Then, utilizing simulations to arbitrary degrees of precision, the threshold value of $b$ may be calculated numerically.

{An} average star of $1.3 M_{\odot}$ {has} a main sequence {lifespan} of around  $\eta\sim 6\times 10^9$ yr.
{
In simulations, the star prototype was set at around  $1.3\, M_{\odot}$. 
 A "habstar," or a star that is exceptionally hospitable to an Earth-like planet \citep{10.1086/345779}, is one way of describing a solar twin. Variability, mass, age, metallicity, and close companions are all factors examined while evaluating a habstar. The stipulation that the star {is} on the main sequence for at least 3 Gyr imposes an upper limit of  $1.5\, M_{\odot}$, which corresponds to a hottest spectral type of F1V \citep{10.1086/345779}. As a result, we assume that a star prototype is a viable Solar-like star with a mass that is closer to Copernican principle criteria and corresponds to the upper limit for habitable Solar-like systems \citep[ i.e., $1.3\, M_{\odot}$][]{10.1017/S1473550410000042}.}

The time step of the evolution of CA is then  {specified}  by $\eta$ and $\mathcal{T}$,
{as follows} $\eta/\mathcal{T}=6\times 10^{9}/500\sim 12\times10^6$yr.
To compute an evolutionary sequence, we {create} an initial CA state by populating cells
with stellar systems based on Eq. \ref{eq:quadratic}. We {next} infect 100 systems within the patch at random with life.
This defines the lattice configuration at the initial time.




{The simulations were run on the SUPERAST,  HPE ProLiant DL380 Gen10 2x Xeon 4210-S 64GB SFF  server located at the Department of Astronomy, Faculty of Mathematics, University of Belgrade \citep[description in][]{Kov21}. There are two computational nodes in SUPERAST. Each node has 40 cores, 128GB RAM,  2TB memory on SSD, and 3 DP GFLOPS (DoublePrecision Giga Floating Point Operations Per Second). We  developed the Python code for simulations and data presentations.}

\section{Results}

{The simulation results are presented in this section to assess the following topics: i) characterizing general spatial patterns of feasible transfers; ii) detecting emergent characteristics in the phase space given by simulation parameters.
\indent For the simulation, we employed a CA with $N=10000 $ cells dispersed in a  $100\times 100$  grid.
 The dynamics of SIR equations should, in theory, be insensitive to {the} population size N \citep{10.1063/5.0028972}. {Thus}, for example, the spatio-temporal patterns of life spreading are not alternated {by} employing the CA with $200\times 200$ cells. 
{On the other hand}, complex models will assume {particular} extra effects that alternate life-spreading properties locally (e.g., viable material speed) in such a way that invariance is destroyed \citep{10.2139/ssrn.3625361}.}

\subsection{Spatial patterns}

{Identifying observable spatial patterns is one of the {critical} problems about the propagation of viable material within the galactic patch. There are two types of spatial patterns that might be explored in this context.    As predicted by \cite{10.1088/2041-8205/810/1/l3} stationary patterns would remain unaltered across time with intermittent zones of  {a} high density of life-bearing hosts. {On the other hand}, we show here that spatio-temporal patterns that change over time might {propagate} like oscillatory waves or even turbulence.  }

\begin{figure*}[h]
\centering
\begin{tabular}{cc}
\subfloat[]{\includegraphics[width=2.5in]
{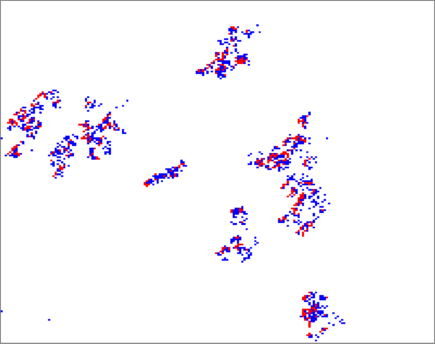}}&
\label{a}
\subfloat[]{\includegraphics[width=2.5in]
{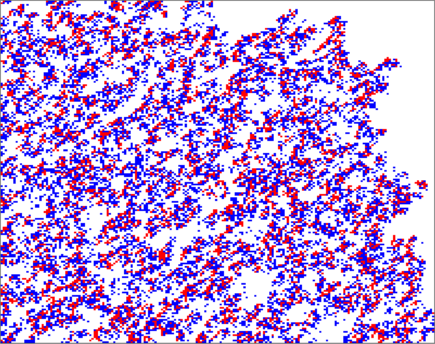}}\\
\label{b}
\subfloat[]{\includegraphics[width=2.5in]
{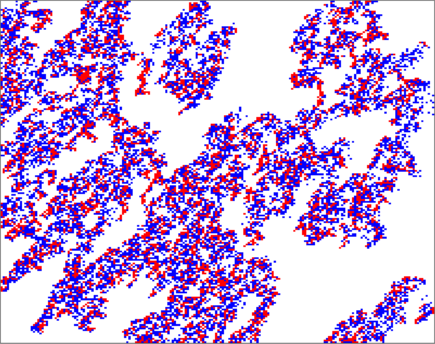}}&
\label{c}
\subfloat[]{\includegraphics[width=2.5in]
{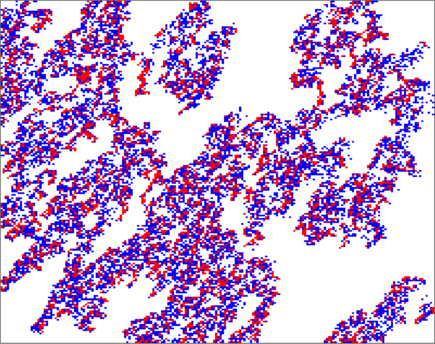}}\\
\label{d}
\subfloat[]{\includegraphics[width=2.5in]
{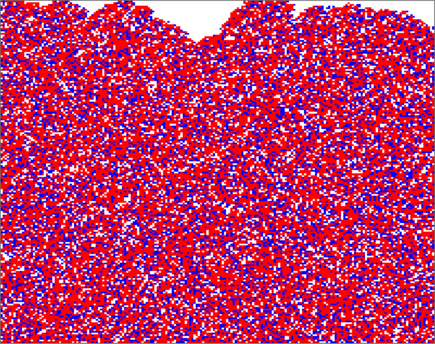}}&
\label{e}
\subfloat[]{\includegraphics[width=2.5in]
{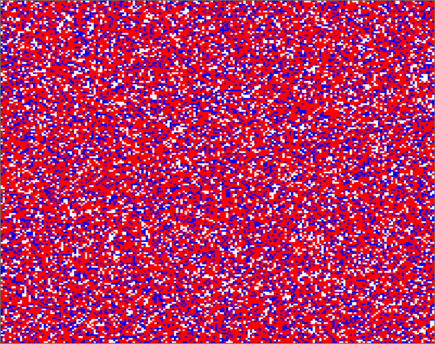}}\\
\label{f}
\end{tabular}
\caption{Spatial patterns in the life-diffusion network for various combinations of parameters when the distribution of hosts is  homogeneous. Colors denote: white- homogeneous distribution of {habitable} hosts;  red is life-bearing hosts and blue cells are life-terminated hosts. Upper row:  Localized life outbreaks are self-limiting in size   for $\beta= 0.62,$ at time instant $t=4.6$(subplot a) but as time proceeds life-islands are still self-limited in size $t=17$ (subplot b).
        Middle: appearance of longer and narrow chains (waves) of life-outbreaks for  $\beta= 1.2$, at time instant $t=16$ (subplot c), with time waves have tendency to connect  $t=21$  (subplot d)
        Bottom:Stable spiral waves of life outbreaks (filaments) for $\beta=2.75$ at $ t=16$ are broad (subplot e) and do not  break
        with time  $t=20$ (subplot f).}
\label{fig:homog}
\end{figure*}

If hosts (e.g., planetary systems) are continuously distributed, life-dispersal spatial patterns can be best seen. {Therefore}, we assumed 100 randomly {inhabited} systems  in such a grid.
In Fig. \ref{fig:homog} we exhibit several evolutionary sequences of continuously distributed hosts affected with life. 
The emerging states are formed either by an entanglement of clusters, 'turbulences' or  'spiral waves,' characterizing the dynamics of the emergence of systems either infected with life, habitable, or sterilized.

{The parameter $beta$, which determines the life spreading rate $b$, affects the spatial patterns.}
When the condition  $\beta=0.62)$ is met, the habitable (white) and inhabited (red) planets coexist in self-organized clusters of restricted size (Fig. \ref{fig:homog},  top left subplot).
It {indicates} that isolated life outbreaks can fill in the lattice (right top subplot), but they can {no longer expand}; thus, they {diminish and eventually vanish}.
{When the} $\beta$ parameter {is} increased, the filaments of life-hosting structures grow in size and develop moving patterns (see 'turbulence' in the middle subplots and 'spirals' in the bottom subplots). The {effectiveness} of filling in the grid is the difference between two types of waves.  Spirals are more effective in filling in the grid {so that} for a short period, {they} can populate the whole grid (bottom subplots).
The interplay of the introduced CA rules sensitively determines whether life-bearing filaments can coexist on the grid or not. 
{Whether or not} life-bearing filaments may live on the grid is determined by the interaction of the introduced CA rules.
{
We will go into more detail about stationary and moving patterns in the {following} subsection, focusing on conditions within phase space that allow for global extinction and persistence of life transmission as stationary patterns and transition between extinction and persistence as a moving spatio-temporal pattern.}

\subsection{Phase space of life transmission}

{In this part, we {look at} two different aspects of stationary and moving life-transmission patterns: i) {the} features of life transmission phase spaces specified by model parameters $(b,g)$ and $(g,T)$; ii) the time it takes for life transmission to end in the galactic patch, which provides information regarding life spreading stability.

\subsubsection{Extinction, persistence and transition}
We are particularly interested in determining the phase transition between global termination and persistence of the life-spreading process, taking into account the distribution of planetary systems as described by Eq. \ref {eq:quadratic}. The global termination and persistence of the life-spreading process are both expressed as static spatial patterns. {However}, the transition between these two states has {a} spatiotemporal dynamic.}

\begin{figure}
    \includegraphics[width=0.5\textwidth]{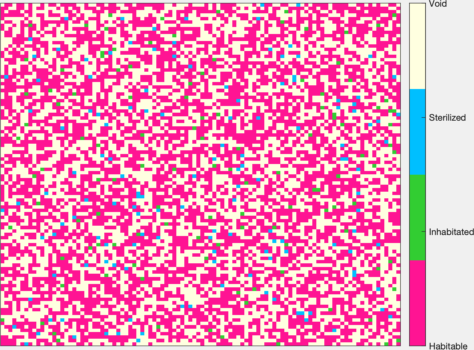}
    \caption{CA with   initial configuration of stellar systems sampled from density distribution given in   Eq. \ref {eq:quadratic}.
        Red, green, blue, and yellow stand for habitable, inhabited, sterilized planetary systems and voids between systems. }
    \label{fig:CA}
\end{figure}

{Figure \ref{fig:CA}} shows a random realization of an initial lattice. Due to the presence of voids, the dynamics of spreading are {not obvious} as they are in idealized {circumstances} (compare to  Fig. \ref{fig:homog}).

\begin{figure*}[!h]
\subfloat[]{\includegraphics[width=2.8in]{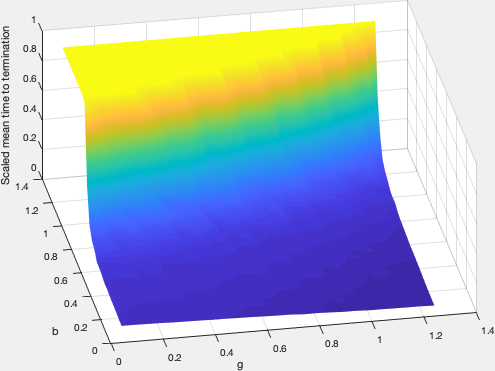}}
\label{a}
\hfil
\subfloat[]{\includegraphics[width=2.8in]{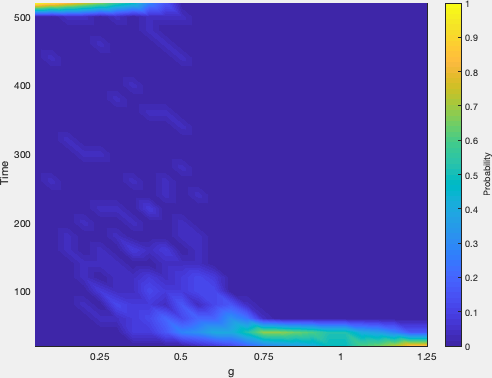}}
\label{b}

\caption{Average simulation realization of (b,g) phase space from 10  runs when the rate of life recovery after sterilization is 0.25. \textit{Left}: The blue color defines the termination part of phase space, and other colors stand for perseverance subregion. The yellow color denotes that life spreading continues during the whole period of simulation. Time is averaged and scaled. \textit{Right}: The probability that the life spreading has gone extinct after a waiting time $t $ when the g parameter is varied. Time is averaged but not scaled. }
\label{fig:int1}
\end{figure*}

{The} phase transition between the persistence and extinction of life diffusion within the Galactic patch is determined via simulations. The {interaction} of sterilization and life infection rates defines this phase space (g,b).
The perseverance part of phase space is defined {as the region} in which the spread of life will continue for the whole simulation run $\mathcal{T}$. 
{The section} of phase space where the life dispersal will terminate at {any} time instant  $t<<\mathcal{T}$ is called the termination part.
The realizations of (b,g) parameters space were calculated by simulating 100 $\times$ 100 cells {lattice} with 10 independent runnings.  {With} time step 1, each run lasted 500 random time units. The lattice {was} initially set with a randomly chosen sample of planetary systems from a density distribution  (Eq. \ref{eq:quadratic}) at the start of each session. Then 100 planetary systems were chosen at random to be life-bearing. 

Both $\beta$ and $g_0$, {which were used to} calculate the (b,g)-plane, {have} values between 0.05 and 1.25, with an increase of $\Delta \beta=\Delta g_0=0.05$. After sterilization, the rate of life recovery {is set} to 0.25.
We {estimated} a joint 2D probability density in (b,g) phase space by averaging the results of all 10 independent runs  (see Fig \ref{fig:int1}). The extinction zone (blue) is large, but the perseverance zone (yellow) is smaller and triangular {in} shape (left plot).

{For a} range of g between 0.5 and 1.25, typical waiting time (T) is often brief until life is terminated everywhere on the grid (blue region left plot).  
{However}, in $(g, T)$ space, we discovered that the life spreading  {persists} only for a limited range $g<0.4$ throughout the simulation (right plot). The likelihood of life spreading for a short time, on the other hand, is higher.

{With a {low} probability}, life can spread throughout intermediate periods (white dots in blue region).
{There are also} phase space transition zones between extinction (blue region) and perseverance (yellow region), which are colored with green and orange (left plot).
Moreover, transition areas can occur with very low probability (left plot  Fig.  \ref{fig:int1}), {which size is regulated} by parameter b.
 
\begin{figure*}[!h]
\subfloat[]{\includegraphics[width=3in]{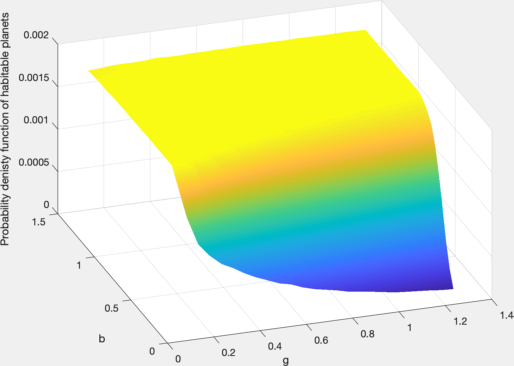}}
\label{a}
\hfil
\subfloat[]{\includegraphics[width=3in]{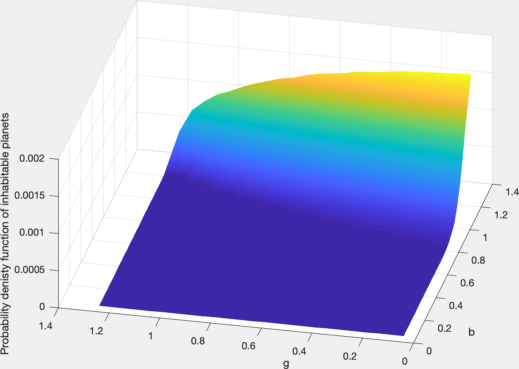}}
\label{b}
\subfloat[]{\includegraphics[width=3in]{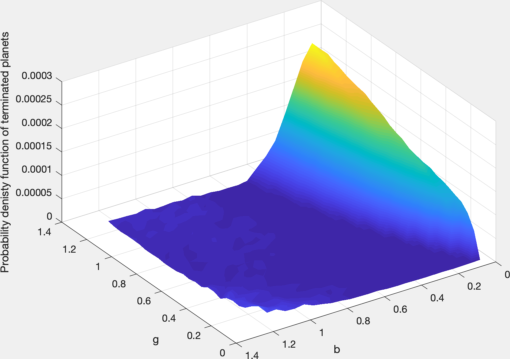}}
\label{c}
\caption{Probability density functions of habitable, inhabited, and life-terminated planetary systems for simulation in Fig. \ref{fig:int1}. a) habitable hosts occupy almost whole phase space, except a tiny band for small b and larger g; b) inhabited hosts are more probable to occur  in the upper plateau (see  right plot in Fig. \ref{fig:int1});c) life-terminated hosts are most likely to occur in lower 
plateau (compare to right plot in Fig. \ref{fig:int1}).}
\label{fig:int11}
\end{figure*}

{The} probability density function of the occurrence of habitable, inhabitable, and life terminated hosts (planetary systems) for simulations presented in Fig. \ref{fig:int1} {is summarized in} Fig. \ref{fig:int11}.

{For} parameter ranges $b<0.5$ and $g>0.45$, habitable hosts can occupy {almost} the whole phase space. Inhabited systems are more likely to exist in the zone specified by overall parameter $b$ values except for $g>1$, which is seen as the upper yellow region in the central plot of  Fig. \ref{fig:int11}.
 When the parameter $b$ is smallest and $g>0.7$, life aborted hosts are most likely to occur near the bottom plateau, which is depicted as the yellow region in the right plot in Fig. \ref{fig:int11}.

\begin{figure*}[!h]
\subfloat[]{\includegraphics[width=3in]{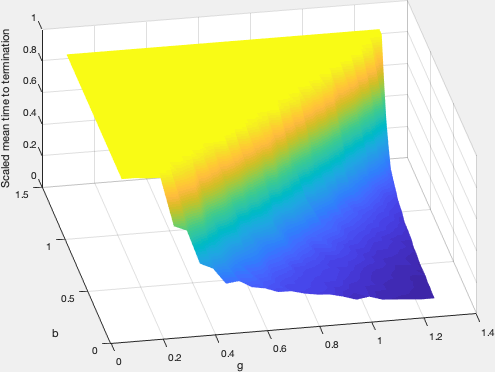}}
\label{a}
\hfil
\subfloat[]{\includegraphics[width=3in]{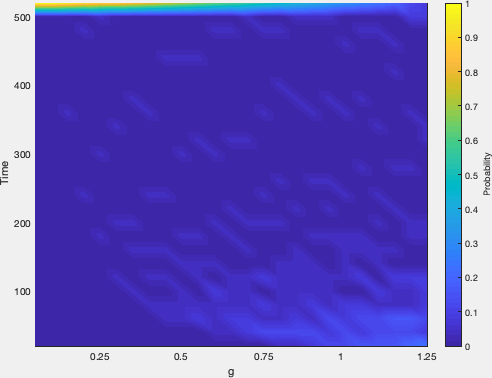}}

\label{b}

\caption{The same as Fig. \ref{fig:int1} but for the high rate of life recovery $0.5$. Left plot: The blue color defines the termination part of phase space, and other colors stand for perseverance subregion. The yellow color denotes that life-spreading continued during the whole period of simulation. Time is averaged and scaled. The right plot is the probability that the life spreading has gone extinct after a waiting time $t $ when the g parameter is varied, and the life recovery rate is high $\sim 0.5$.  Time is averaged but not scaled.}
\label{fig:int111}
\end{figure*}

{The} perseverance area of phase space (yellow) and transition section of phase space (green and orange) are expanded when the rate of life recovery is high $0.5$ (see the left plot in Fig \ref{fig:int111}). As a result, the typical waiting time T until life ends is usually {significant}.

The probability of life termination {occurring} early in simulation is extremely low (see the right plot in  Fig \ref{fig:int111}), as seen by the disappearance of the bottom plateau.

\begin{figure*}[!h]
\subfloat[]{\includegraphics[width=3in]{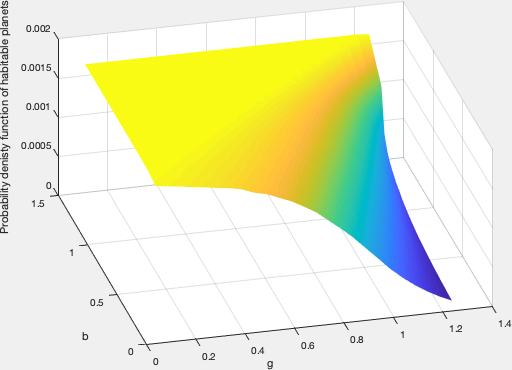}}
\label{a}
\hfil
\subfloat[]{\includegraphics[width=3in]{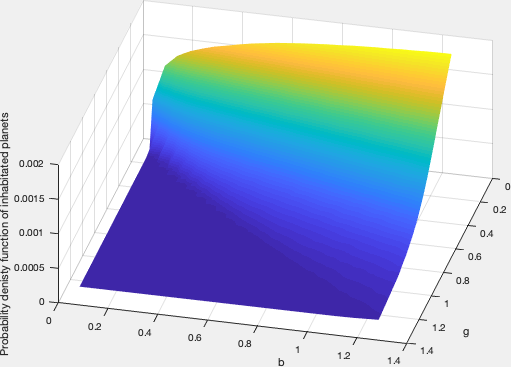}}
\label{b}
\subfloat[]{\includegraphics[width=3in]{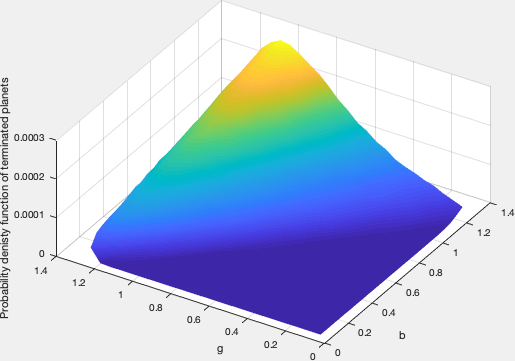}}

\label{c}

\caption{Probability density of habitable, inhabited and terminated systems  for simulation in Fig. \ref{fig:int111}. 
From left to right: habitable systems occupy larger than a half of phase space; inhabitable systems are more probable to occur in the upper plateau (see {the} right plot in Fig. \ref{fig:int111}); terminated hosts are most likely bellow the upper plateau  (compare to right plot in Fig. \ref{fig:int111}).}
\label{fig:int15}
\end{figure*}

The probability density functions of occurrence of habitable, inhabitable, and life terminated hosts are given in Fig \ref{fig:int15} for large values ($\sim0.5$) of recovery rate of life. There are noticeable differences between Fig. \ref{fig:int11}, and  Fig \ref{fig:int15}.
{Habitable} planets occur with a high probability (left panel) in more than half of phase space. {However}, the probability density space of inhabited hosts (middle panel) is slightly larger than for $b<0.5$. A life-terminated system's probability density region is also slightly larger (right panel).

\subsubsection{On the stability of life spreading}
Through the simulations, when the parameter b is larger, the waiting period $T$ before life transmission is stopped can be quite long, but when $b< 0.5$, the waiting time is regulated by $g$ (see Fig. \ref{fig:int1} and Fig. \ref{fig:int111}). This shows that the waiting time $T$  {dependence} on $g$ is something to ponder carefully.
When $b<0.5$, the usual waiting time strongly increases as the ratio $T/g$ approaches infinity in the asymptotic limit of small g.
We can consider this arrangement to be stable for life spreading across the galactic patch. 
If life recovery parameter $b$ is smaller, the probability is high, as seen as an upper plateau, whose reflection in $(b,g)$ phase space is shorter (see right plots in Fig. \ref{fig:int1} and compare it to Fig.  \ref{fig:int111} ).
{When} $T/g \rightarrow O(1)$ (i.e. the ratio approaches a finite non-zero value, see right plot in Fig. \ref{fig:int1}) a lower plateau {appears}, which can vanish completely when $b\geq0.5$ (see right plot in Fig. \ref{fig:int111}).
It could be interpreted as a circumstance in which the life-spreading is unstable, with a decreased probability.
The third region with close to zero probabilities can occur anywhere between these two plateaus and is influenced by the $b$ parameter (see left column of plots  Fig. \ref{fig:int1} and  \ref{fig:int111}).

\section{Discussion}
Within a simulated Galactic patch, we discovered at least two probable spatial patterns of life transmission, corresponding either to stationary cluster patterns proposed by \cite{10.1088/2041-8205/810/1/l3} or moving spatio-temporal patterns (in the form of waves or turbulences).
The 'speed' with which the spatio-temporal patterns fill the spatial patch varies. The value of parameter $\beta$, which determines the life-infection rate b, affects the diverse life-spreading spatial patterns. {As} $ \ beta$ increases, {the} spatial forms of life transmission develop in size and form moving patterns of various speeds.
The sterilization and life infection rates (g,b) were used to create a phase space of life transmission.
The CA simulations indicate the partitioning of phase space into two subregions: the perseverance subregion, where life diffusion will continue for $t=\mathcal{T}$, and the termination subregion, where life diffusion will come to a halt at substantially {little} time instants $t<<\mathcal{T}$.
 The third transition subregion is related to time instants progressively approaching the total simulation time $\mathcal{T}$.  
These phase space partitions are defined by the parameters b and g, as indicated. Larger values of b, for example, dramatically expand the phase space region where life can spread.

This statement is interesting, as it seems to be the opposite of what one would expect.  {In a given patch, the origin of life can be intrinsic abiogenesis or the “infection” of the patch by panspermia. In either case, its probability is the site's inherent property and cannot {relate to} the independent probability of a catastrophe. For example, the life-terminating events could be non-local, e.g., SNe or gamma-ray bursts, affecting multiple habitable sites. In fact, by allowing greater flexibility, one can speculate that effects similar to those in the context of a metapopulation in biology \citep{10.1016/B978-0-12-384719-5.00172-6} can take place. If the rate of extinction {exceeds} the rate {of} recolonization, the metapopulation {as a whole will go} extinct in the galactic patch, and vice versa. Even if recolonization exceeds the extinction rate, there is a risk that metapopulations with only a few local populations will all go extinct {simultaneously}, resulting in the metapopulation's extinction \citep{10.1016/B978-0-12-384719-5.00172-6}. At any given period, catastrophic events in space {must hit irregularly enough to wipe out} only a part of local populations in the galactic patch, or the metapopulation as a whole could perish \citep{10.1016/B978-0-12-384719-5.00172-6}. For example, {environmental disturbance agents} on our planet, such as fires and storms, are known to distribute death unevenly in space, despite their ability to {affect} a huge area in a short  time \citep{10.1016/B978-0-12-384719-5.00172-6}.
Still, as a result of climate change, catastrophic events are affecting much broader areas. {Galactic patches may} include two types of environment-disturbing agents: those that strike randomly and those that strike systematically.} However, based on our simulations, it appears that the persistence of life spreading is {impacted by} the life-infection rate in irregular disasters.

{For} the self-formation of spatial patterns, the discrete character of the hosts involved in life spreading mechanisms, as well as the transmission process rate, are critical.

{We also observe that life transmission is sensitive to the rate of life recovery.
The perseverance and transition parts of phase space are extended when the life recovery rate is large (0.5). {This is because} the time it takes for life to end in the galactic region is usually significant. {On the other hand} transition zones (i.e., moving spatial patterns) can occur with very low probability when the rate of life recovery is {lower} (0.25). {This} parameter may vary from host to host, but we accounted for it as a constant for simplicity's sake.
Because large-scale systems can have varying couplings among their components  \citep{10.1103/PhysRevLett.107.124101}, life-transfer observables on long-range temporal and spatial scales within the Galactic region are expected to be less resolvable than those on smaller scales, such as life spreading within a single planetary system.}

{
The waiting time's dependency on g appears to be critical {to measure} the stability of life spreading mechanisms inside the patch under consideration.
The ratio $T/g$ {is likely to approach} an infinite or finite non-zero value. In the latter {instance}, there is a risk of life termination over the region due to instabilities in life transmission.
The former could be interpreted as a scenario in which the spread of life is stable.
{Inter-species} interactions such as competition, predation, and symbiosis can define biodiversity within a region {throughout} life-spreading processes. {However,} because they can join to form a {more extensive} and  complicated network {of connections}, these are not the only sorts of interactions.
Species, on the other hand, can minimize competitive interaction and improve the chances of future colonization. Species can, for example, use alternate life dispersion methods that are usually encouraged by natural selection. This could be influenced by physical occurrences that provide {an} opportunity for non-competitive dispersal.}
The patterns may develop in size and eventually outgrow the system of natural hosts (i.e., planetary systems or planets) when the parameter b increases, as shown in the middle panel of Fig. \ref{fig:homog}. In the parameter space of larger values of b, life-bearing hosts may colonize another {patch} (see bottom panel of Fig. \ref{fig:homog}).
We can not expect patches to be completely separated from one another if we assume the latter scenario.
{
Finally, because life diffusion is disseminated {mainly} through vectors, the variables are naturally discontinuous and heterogeneously distributed.
Larger and more sophisticated multi-patches settings will be the focus of our future research. We expect each patch to evolve separately through successive and varied forms of biotic forcing (colonization ability, reproduction, growth efficiency,  dispersal capacity, and other life strategies) according to the patch-mosaic model in biology.  These processes promote the coexistence of a large number of species on much larger spatio-temporal scales  \citep [see][]{10.1371/journal.pone.0018912}.}

{High} levels of isolation, according to {the} biogeography of our planet, {discriminate} against dispersal. The loss of dispersal ability on oceanic islands, such as flightless birds  \citep{Carlquist}, is an example of this.

Organisms in unique isolated habitats as subterranean caves, mountaintops, deep-sea trenches, thermal
vents, and hot springs pose reduced dispersal characteristics.  \cite{Darwin}  explained this phenomenon
by counteracting selective forces acting before and after
colonization of a remote island.  
{Because of} high-level isolation, species can lose their ability to disperse in just a few generations.
Based on this, we can speculate that life forms on other planets can be either interplanetary dispersal-prone or -against depending on their host planet isolation (or level of possible material exchange in that region).
We can hypothesize possible categories of interplanetary dispersal barriers based on {various} stresses organisms must endure:  physiological, environmental, or behavioral, environmental hazards, or behavioral difficulties in the environments they traverse. 

{In addition}, dispersal barrier traversability can be classed as corridors, filters, or sweepstakes, based on the complete range of probabilities of passing a barrier, from extremely probable to exceedingly improbable, respectively.\citep{Simp65}.
{In densely} packed planetary systems, such as the TRAPPIST-1 system, corridors may be formed. {For example}, three of seven planets are located within the HZ and have rocky compositions \citep{Gillon16, Gillon17}. 
Estimated orbital periods on the order of days, {as well as} orbital separations of $<0.01$  AU, {speed up} material {transfer} between these planets, {perhaps} four to five times faster than between Earth and Mars \citep{10.1073/pnas.1703517114, Krijt17}.
{For} solar-like planetary systems and cosmic regions {akin} to the solar neighborhood, filters and sweepstakes (extremely improbable transmission pathways) would be { beneficial}.

On our planet, life recovery {happens} after each mass extinction as correlated events \citep{10.1098/rspb.1999.0773, NP03}, but with non-uniform 'speed'.
{One can imagine a} context in which violent events in synergy with spatial patterns in Galactic patches can induce selection for 'the rate' of life recovery and spread, similar to what we know about different speeds of diversity and complexity recovery after the Permian-Triassic (PT) catastrophe \citep{10.1126/sciadv.aat5091}. {Violent global regulatory mechanisms can be supernova explosions that have been first discussed as such by  \citet{10.1017/S1473550410000042}, while local ones comprise meteor impacts, volcanism, and chemical compositions.  For instance, according to} \cite{10.1098/rspb.1999.0773} extinctions on Earth at { various} times are correlated, and extinctions within a single stage are not independent events.
\cite{NP03} examined the number of families of marine animals that appeared with the number that went extinct in each geological stage since the Phanerozoic began and found that there are origination peaks that match all of the significant extinction peaks. However, there is {no perfect} correlation between the two curves characterizing these two sorts of events.
{Moreover}, on our planet, different aspects of life recovery do not occur at the same time. {There is a distinction to be made between diversity  and complexity recovery.}
For example, nearly all life on Earth was wiped around 252 million years ago during Earth's largest mass extinction \citep{Sep84},  known as the PT mass extinction, named after the two geologic periods it delineates.
The recovery of marine ecosystems is thought to have taken several million years \citep{10.1016/S0169-5347(98)01436-0}, with gradual recovery from the bottom to the top trophic levels  \citep{10.1038/ngeo1475}. However, data {imply} \citep{10.1126/sciadv.aat5091} that the restoration of the marine ecosystem  ($\sim 50$ Myr) {took an} order of magnitude {longer} than the recovery of biological diversity($\sim$ 5 Myr). 

{In the same way that a synergy of mechanisms controls the evolution of life forms on our planet}, a synergy of mechanisms may exist {to govern the} evolution of life forms that cause relatively moderate but frequent life recovery and spread after each reset of life on a planet or planetary system.
{Individual} life forms migrate over space, which is a {critical} element of Earth's ecosystems. {Migration} may also be a key competitor with local planetary interactions within the Galactic patch, thereby influencing species preservation and biodiversity.

The parameters $b$ and $g$ could be generalized functions that include interplanetary system covariates such {as} observable biological, chemical, or other environmental characteristics. It is also {feasible} to add data, either in a static sense, where the distance kernel may allow for material exchange direction or in a dynamic sense, where the kernel could alter over time as new exoplanet data becomes available. The phase space of life-spreading can thus be recreated using the biomarkers and biosignatures data from a specific Galaxy patch.
{It is} critical to understand that these spatially heterogeneous reactions might modify transmission patterns in unexpected ways.
{As a result,} in order to qualitatively explain these findings, thorough knowledge is still required, which could lead to additional research.

{According to} the recent study by \cite{Totani20} the predicted number of abiogenesis events for a star, galaxy, or perhaps the entire observable universe is  $< 1$. 
{Because} the volume of the observable universe is less than $< 10^{-78}$ of the total inflation universe, \cite{Totani20} {suggested } that it is impossible to predict more than one abiogenesis event in such a small location without sufficient evidence.
{Even if} Earth were the only planet with life within the observable universe, life could nonetheless emerge on innumerable planets throughout the inflationary cosmos.
If abiogenesis is the only process for the {origination} of life {everywhere},  \citet{Totani20} calculations demonstrate that 
{detecting non-terrestrial biosignatures in Solar and extraterrestrial systems will be extremely improbable}.

difficult.

{Totani's analysis has some strong counterarguments. { First,} although the early Big Bang's inflation episode provides elegant solutions to some observable universe dilemmas, it has significant difficulty explaining how the inflation process began and ended. \citet{10.1016/j.physletb.2013.05.023,10.1016/j.physletb.2014.07.012} and \citet{10.1088/0264-9381/33/4/044001} demonstrated that Planck satellite data ruled out the simplest inflation models and that the remaining inflation models necessitate more parameters, fine-tuning, and more improbable initial conditions. At the very least, it appears that the inflation theory will need to be extensively revised, while numerous suggested changes do not appear to {impact} Totani's main conclusion. Second, Totani's research is limited to carbon-based biology (RNA-based biology). However, while other elements may be used to create live organisms, carbon-based biochemistry is widely acknowledged as the most ideal for { producing} complex molecules, which is essential for any possible kind of living or sentient beings.
Finally, while Totani's analysis is based on current research, it is still vulnerable to future laboratory tests of the RNA world hypothesis. If extraterrestrial species of a different origin than those on Earth are discovered in the future, it {will} imply an unknown mechanism at work to polymerize nucleotides considerably quicker than random statistical processes, as Totani himself points out.}

However, if interplanetary and (or) interstellar transmission of viable material is allowed \citep{Nicholson09, 10.1007/s11214-010-9671-x} life detection elsewhere is more likely to occur. {In seminal works such as \citet{Zuck85} and \citet{10.1016/j.newast.2006.04.003} various plausible motivations for life migrations have been found. }
 
Nonetheless, certain violent events may indirectly improve life transmission.
{The} presence of the supermassive black hole at the Galactic core, for example, might result in a large flux of ionizing radiation, causing planetary atmospheric erosion and {possibly} severe consequences on surface life \citep{10.1051/0004-6361/201834655}. At the same time, erosion effects could have increased the number of terrestrial planets formed from a gaseous envelope of sub-Neptune planets \citep{10.3847/2041-8213/aaab46}, or {stimulated} prebiotic chemistry or even photosynthetic activity on planets {without} incoming radiation  \citep{10.3847/1538-4357/ab1b2f}.

If life can be transmitted across stellar systems, the highest rate of catastrophic events could be {mitigated} by the {possibility} that life can relocate swiftly to safer areas  \citep{10.3390/life10080132} or that life recovery {rate} is high within denser Galactic regions like the bulge.

{The phase volume within which life diffusion occurs is the multidimensional space, determined by mass--viable transfer--extinction interactions, wherein life can exist. This phase space represents an emergent property informed by all levels of interactions and the laws of thermodynamics. Indeed, { specific} {neighborhood} systems  are just one point within this space. The phase space, as we {have} seen,  can increase or decrease depending on assumed interactions.  We believe there is a concentration gradient, {so the} material will predominantly move from high to low concentration regions in the galactic patch. The phase space itself { may} have {its} gradient defined by the ability of life diffusion to overcome the threshold imposed by parameter $b$, leading to an increase or a decline in diversity.}

\section{Conclusion}
{The availability of biosignatures from planned ground and space surveys will allow precise measurement of phase-space as the "standard" for understanding life-spreading mechanisms.}

{According to our results}, we identify three qualitatively different phase space regions in terms of possible interstellar life spreading outcomes based on transmission and sterilization parameters: perseverance, extinction, and transition zone.
{Based on} our simulations, the size of these three regions is apparently governed by infection rate (parameter $b$). {Thus,} three possible geometrical shapes ('waves') of interstellar life-spreading have been identified.
Finally, our results suggest the sensitivity of interstellar life-transmission processes by revealing how small changes in the parameters can cause changes in population-level outcomes such as the probability density of inhabited and sterilized hosts (planetary systems).
Though proposing a generative mechanism for life-transmission is beyond the scope of this research, our work brings a new piece to the puzzle of (un)correlated life-origin.

\ack[Acknowledgement]{The Reviewers have provided extremely detailed and useful suggestions regarding the text clarity and presentation, for which we are grateful. The author acknowledge funding provided by the Faculty of Mathematics University of Belgrade  (the contract 451-03-68/2020-14/200104),  through the grants by the Ministry of Education, Science, and Technological Development of the Republic of Serbia. A. K.  acknowledges the support by  Chinese Academy of Sciences President's International Fellowship Initiative (PIFI) for visiting scientist. A part of this work has been presented as a short oral presentation  at the Europlanet Science Congress (Virtual meeting) 2020.}




\end{document}